\documentclass[prl,twocolumn,showpacs,preprintnumbers,amsmath,amssymb]{revtex4-1}
\usepackage{graphicx}

\begin{document}
\title{Dependence of carrier doping on the impurity potential in
    transition-metal-substituted FeAs-based superconductors}%

\author{S. Ideta$^{1}$, T. Yoshida$^{1,2}$, I. Nishi${^1}$, A. Fujimori$^{1,2}$, Y. Kotani$^{3}$, K. Ono$^{3}$, Y. Nakashima$^{4}$, S. Yamaichi$^{4}$, T. Sasagawa$^{4}$, M. Nakajima$^{1,2,5}$, K. Kihou$^{2, 5}$, Y. Tomioka$^{2,5}$, C. H. Lee$^{2,5}$, A. Iyo$^{2,5}$, H. Eisaki$^{2,5}$, T. Ito$^{2,5}$, S. Uchida$^{1,2}$ and R. Arita$^{2, 6}$
}
\affiliation{$^1$ Department of Physics, University of Tokyo, Bunkyo-ku, Tokyo 113-0033, Japan\\
$^2$ JST, Transformative Research-Project on Iron Pnictides (TRIP), Chiyoda, Tokyo 102-0075, Japan\\
$^3$ KEK, Photon Factory, Tsukuba, Ibaraki 305-0801, Japan\\
$^4$ Materials and Structures Laboratory, Tokyo Institute of Technology, Yokohama, Kanagawa 226-8503, Japan\\
$^5$ National Institute of Advanced Industrial Science and Technology, Tsukuba 305-8568, Japan\\
$^6$ Department of Applied Physics, University of Tokyo, Bunkyo-ku, Tokyo 113-8656}
\date{\today}%
\begin{abstract}
In order to examine to what extent the rigid-band-like electron doping scenario is applicable to the transition metal-substituted Fe-based superconductors, we have performed angle-resolved photoemission spectroscopy studies of Ba(Fe$_{1-x}$Ni$_{x}$)$_2$As$_2$ (Ni-122) and Ba(Fe$_{1-x}$Cu$_{x}$)$_2$As$_2$ (Cu-122), and compared the results with Ba(Fe$_{1-x}$Co$_x$)$_2$As$_2$ (Co-122). We find that Ni 3$\it{d}$-derived features are formed below the Fe 3$\it{d}$ band and that Cu 3$\it{d}$-derived ones further below it. The electron and hole Fermi surface (FS) volumes are found to increase and decrease with substitution, respectively, qualitatively consistent with the rigid-band model. However, the total extra electron number estimated from the FS volumes (the total electron FS volume minus the total hole FS volume) is found to decrease in going from Co-, Ni-, to Cu-122 for a fixed nominal extra electron number, that is, the number of electrons that participate in the formation of FS decreases with increasing impurity potential. We find that the N$\acute{\rm{e}}$el temperature $T_{\rm{N}}$ and the critical temperature $T_{\it{c}}$ maximum are determined by the FS volumes rather than the nominal extra electron concentration nor the substituted atom concentration.  
\end{abstract}

\pacs{74.25.Jb, 71.18.+y, 74.70.-b, 79.60.-i}% PACS, the Physics and Astronomy
                             % Classification Scheme.
\maketitle
Carrier doping plays an essential role in a variety of correlated electron systems such as the high-$T_c$ cuprates as well as the newly discovered ion-based high-$T_c$ superconductors. For example, one of the parent compounds $A$Fe$_2$As$_2$ ($A$ = Ba, Sr) abbreviated as \textquotedblleft 122\textquotedblright\ is an antiferromagnetic metal but becomes superconducting (SC) by substitution of monovalent metal atoms for divalent Ba or Sr at the $A$ site, which obviously dopes the Fe-As layer with holes \cite{Rotter, Sasmal}. Superconductivity is also realized via partial substitution of transition-metal atoms such as Co, Ni, and Cu for Fe \cite{Sefat, Canfield, Li, Ni}. According to the rigid-band model, the substitution leads to electron doping and the doped electron number in Ba(Fe$_{1-x}$Co$_{x}$)$_2$As$_2$ (Co-122), Ba(Fe$_{1-x}$Ni$_{x}$)$_2$As$_2$ (Ni-122), and Ba(Fe$_{1-x}$Cu$_{x}$)$_2$As$_2$ (Cu-122) is expected to be $\it{x}$, 2$\it{x}$, and 3$\it{x}$, respectively, while it is $\it{x}$ for Co-122. This is a remarkable phenomenon given the fact that the substitution of transition-metal atoms for Cu in the CuO$_2$ plane of the cuprate superconductors quickly kills the superconductivity \cite{Fukuzumi}.

\begin{figure}[t]
\includegraphics[width=9.2cm]{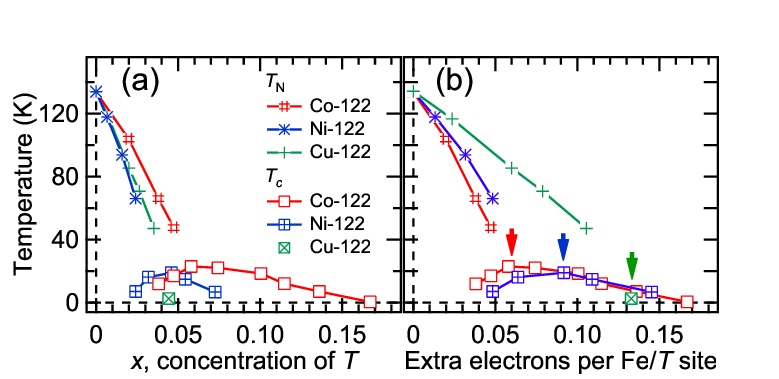}
\caption{(Color online) Superconducting and magnetic transition temperatures ($T_c$ and $T_{\rm{N}}$) of Ba(Fe$_{1-x}T_x$)$_2$As$_2$, ($T$-122: $T$ = Co, Ni, Cu) reported in Ref. \cite{Canfield, Ni}. (a): Plotted as functions of the number of substituted transition-metal atoms per Fe site, $x$. (b): Plotted as functions of nominal extra electron number per Fe site $\it{x}$, 2$\it{x}$, and 3$\it{x}$ for Co-, Ni-, and Cu-122, respectively.}
\label{FigTMBa122_1}
\end{figure}

The phase diagrams of the transition metal-doped BaFe$_2$As$_2$ are reported by Canfield $\it{et\ al}$. \cite{Canfield} and Ni $et\ al$. \cite{Ni}; however, exhibit complex behaviors as shown in Fig. \ref{FigTMBa122_1}, where the SC transition temperatures $T_c$s and the N$\acute{\rm{e}}$el temperatures $T_{\rm{N}}$s of Co-, Ni-, and Cu-122 are plotted as functions of substituted atom concentration $x$ [panel (a)] and nominal extra electron concentration [panel (b)]. The $T_{\it{c}}$ of Co-122 reaches the maximum 24 K at $x \sim$ 0.06. As for Ni-122, the maximum $\it{T_c}$ of $\sim$ 18 K is reached at $x$ $\sim$ 0.05, but the $\it{T_c}$ drops more rapidly with Ni substitution than Co substitution. Cu-122 is largely non-SC except for $x \sim$ 0.044, where very a low $\it{T_c}$ of $\sim$ 2 K has been reported \cite{Canfield, Ni}. The nearly coinciding $T_c$ domes of Co- and Ni-122 in panel(b) (plotted against the nominal extra electron number) suggest that Fermi surfaces (FSs) follow the rigid-band model, but that the small but distinct mismatch of the two $\it{T_c}$ domes suggests that there are some changes in the electronic structure beyond the rigid-band model, that is, the impurity potential of the substituted atoms are likely to influence the electronic structure and hence $T_c$. For Cu-122, the $T_c$ is strongly suppressed, and the $\it{T_c}$ \textquotedblleft{maximum}\textquotedblright occurs at a much higher nominal extra electron concentration of $\sim$ 0.13 than Co- ($\sim$ 0.06) and Ni-122 ($\sim$ 0.10), suggesting that its electronic structure is strongly deviated from that expected from the rigid-band model. Contrary to $T_c$, the drop of the $T_{\rm{N}}$ with doping is determined by the substituted atom concentration $\it{x}$ rather than the number of extra electrons, as seen from the relatively well overlapping $T_{\rm{N}}$ curves in Fig. \ref{FigTMBa122_1}(a) (although the $T_{\rm{N}}$ curve for Co-122 is somewhat deviated from the other curves).

\begin{figure}[t]
\includegraphics[width=9.3cm]{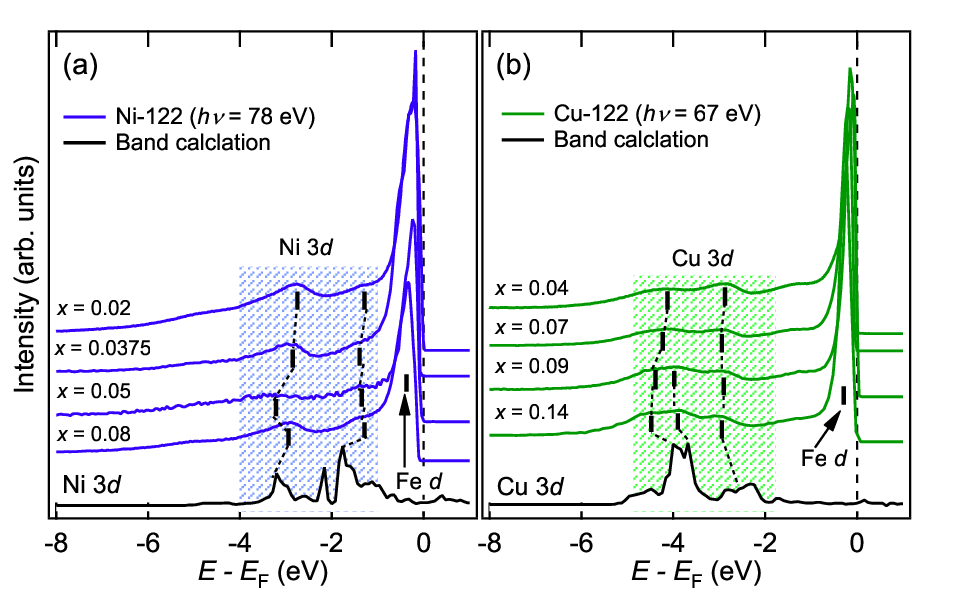}
\caption{(Color online)  Angle-integrated photoemission spectra of Ni-122 [panel (a)] and Cu-122 [panel (b)]. Partial density of states for Ni and Cu 3$d$ orbitals obtained by band-structure calculation \cite{Wadati} are also plotted. Shaded area shows the energy region dominated by Ni and Cu 3$d$-derived states.  }
\label{Valence}
\end{figure}

So far, a large number of angle-resolved photoemission spectroscopy (ARPES) studies have been made on the Co-122. The results have shown that there are two hole FSs and two electron FSs at the two-dimensional Brillouin zone (BZ) center and corner, respectively \cite{Ding, Wray, Sato1, Malaeb, Brouet}. The SC gaps \cite{Terashima, Nakayama} and the three dimensionality \cite{Brouet, Vilmercati, Malaeb} of the hole and electron FSs have also been identified. The FSs evolve following the rigid-band model with electron doping, that is, the volumes of the FSs change according to the number of extra electrons $\it{x}$ and the chemical potential is shifted accordingly \cite{Sekiba, Brouet, Neupane}. However, according to a recent density functional theory (DFT) calculation on supercells \cite{Wadati}, the extra electronic charges are largely distributed on the substituted Co, Ni, or Cu atoms, apparently contradicting with the rigid-band model \cite{Brouet, Chang Liu}. On the other hand, another supercell calculation has indicated that the total electron number enclosed by the FSs changes in proportion to the dopant concentration in Co and Ni-doped LaFeAsO \cite{Konbu}.  

In this Letter, we have performed an ARPES experiment on Ni- and Cu-122, and investigated the evolution of the electronic structure with the strength of impurity potential in going from Co-, Ni-, to Cu-122. We find that the number of doped electrons in Ni- and Cu-122 estimated from the FS volumes is smaller than the value expected from the simple rigid-band model, and decreases in going from Co-, Ni-, to Cu-122 for a fixed nominal doped electron number.

\begin{figure}[t]
\includegraphics[width=9cm]{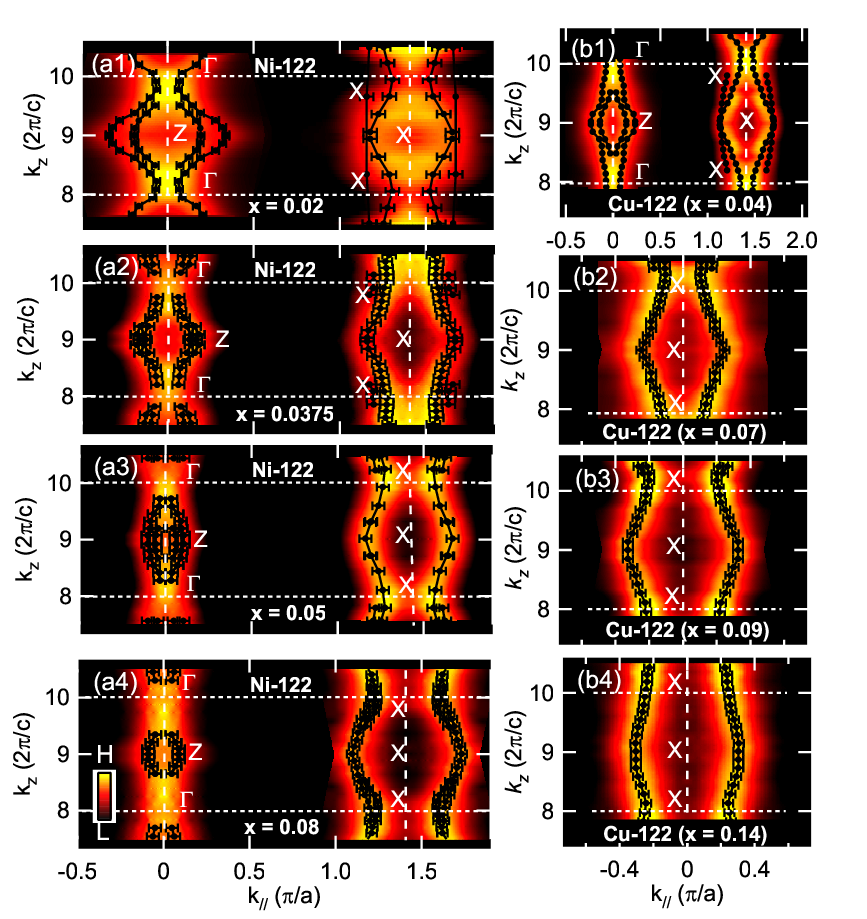}
\caption{(Color online) ARPES intensity plot in $k_{||}$-$k_z$ momentum space for Ni-122 and Cu-122 obtained by $h\nu$-dependent ARPES measurements. (a1)-(a4):  Hole and electron FSs for Ni-122. (b1)-(b4): Hole and electron FSs for Cu-122. $k_{\rm{F}}$ points of the hole and electron FSs are plotted by black dots. The ARPES intensity plots have been symmetrized with respect to symmetry lines.}
\label{FigTMBa122_4}
\end{figure}

High-quality single crystals of Ba(Fe$_{1-x}$Ni$_x$)$_2$As$_2$ with $x$ = 0.02, 0.0375 ($T_c$ $\sim$ 16 K), 0.05 ($T_c$ $\sim$ 18 K), and 0.08 ($T_c$ $\sim$ 5 - 10 K), and Ba(Fe$_{1-x}$Cu$_x$)$_2$As$_2$ with $x$ = 0.04, 0.07, 0.09, and 0.14 were grown by the self-flux method. The Ni and Cu concentrations were determined by X-ray fluorescence analysis and energy dispersive X-ray analysis, respectively. ARPES measurements were carried out at beamline 28A of Photon Factory (PF) using circularly-polarized light ranging from $h\nu$ = 34 to 88 eV and a Scienta SES-2002 analyzer with the total energy resolution of $\sim$ 15 - 25 meV. The crystals were cleaved $in\ situ$ at $T$ = 10 K in an ultra-high vacuum less than 1$\times$10$^{-10}$ Torr. ARPES measurements on Ni-122 ($x$ = 0.02, $T_{\rm{N}} \sim$ 90 K) and Cu-122 ($x$ = 0.04, $T_{\rm{N}} \sim$ 50 K) were performed at temperatures above $T_{\rm{N}}$, $T$ = 110 K and 60 K, respectively. The other samples were measured at $T$ = 10 K.

Figure \ref{Valence} shows angle-integrated photoemission spectra of Ni-122 and Cu-122 in the entire valence-band region. Ni 3$\it{d}$-derived emission is located $\sim$ 2 - 4 eV below the Fermi level ($\it{E_{\rm{F}}}$), and overlaps with the Fe 3$\it{d}$ band. The Cu 3$\it{d}$-derived emission is located at higher binding energies of $\sim$ 4 - 5 eV. This indicates that the impurity potential of the substituted transition metal atoms becomes stronger in going from Co, Ni, to Cu. The spectra are consistent with the density of state given by the super-cell band-structure calculation \cite{Wadati} shown in the same figure.

\begin{figure}[t]
\includegraphics[width=6cm]{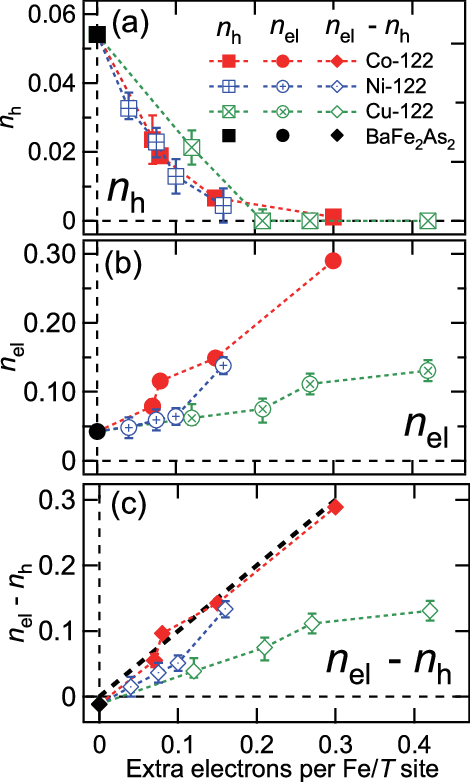}
\caption{(Color online) Hole and electron FS volumes from the ARPES data of BaFe$_2$As$_2$, Co-122, Ni-122, and Cu-122 plotted as functions of the nominal extra electron number. (a): Hole FS volume $n_{\rm{h}}$. (b): Electron FS volume $n_{\rm{el}}$. (c): Total FS volume $n_{\rm{el}}-n_{\rm{h}}$. Black dashed line shows the behavior ($n_{\rm{el}}-n_{\rm{h}}$ = extra electron per Fe/$T$ site) expected from the rigid-band model. BaFe$_2$As$_2$ and Co-122 data are taken from \cite{Brouet, Malaeb}. %(d): Energy shifts of the electron-band minimum and hole-band maximum positions for Co-, Ni-, and Cu-122. The extra electron number per Fe/$T$ site is fixed at 0.16. Data for Co-122 have been taken from \cite{Brouet, Neupane, CLiu}.
%(e), (f):  Shifts of the hole-band maximum around the $\Gamma$ point and the electron-band minimum around the X point. The dashed lines show the shifts expected from the rigid-band model based on the non-magnetic band structure of BaFe$_2$As$_2$ \cite{Ma}. The gray belts indicate the effect of band renormalization by a factor of 1.5 - 2 on the rigid-band model. The data for the parent compound in panel (a) is taken from \cite{Walid_private}.
}
\label{FigTMBa122_5}
\end{figure}

Figure \ref{FigTMBa122_4} shows the summary of FS mapping in $k_{\parallel}$-$k_z$ space for the hole and electron FSs of Ni- and Cu-122. Raw data of hole and electron bands near $E_{\rm{F}}$ are shown in Fig. \ref{FigS2} of Supplemental Material. While the two hole FSs around the Z point are always observed, the hole FS around the $\Gamma$ point shrinks and disappears with increasing Ni concentration, resulting in three-dimensional hole FSs \cite{Singh, Bianconi}. On the other hand, the volumes of the electron FSs around the X point become larger with increasing Ni concentration, indicating electron doping. All the FSs show warping along the $k_z$ direction and the electron FS is less warped than the hole FS. For Ni- and Cu-122, we could not resolve the two electron bands, and therefore, we consider that the two bands are almost degenerate. As a whole, the hole and electron FSs are similar to those of Co-122 \cite{Malaeb}. In the case of Cu-122, the hole FSs were observed only for $x$ = 0.04 as shown in Fig. \ref{FigTMBa122_4}(b1), and we show only the electron FSs for $x$ = 0.07, 0.09, and 0.14 in Figs. \ref{FigTMBa122_4}(b2)-\ref{FigTMBa122_4}(b4) (see Figs. \ref{FigS1} and \ref{FigS2} and the text of \cite{supple}). 

\begin{figure}[t]
  \begin{center}
\leavevmode
    \includegraphics[width=9.2cm]{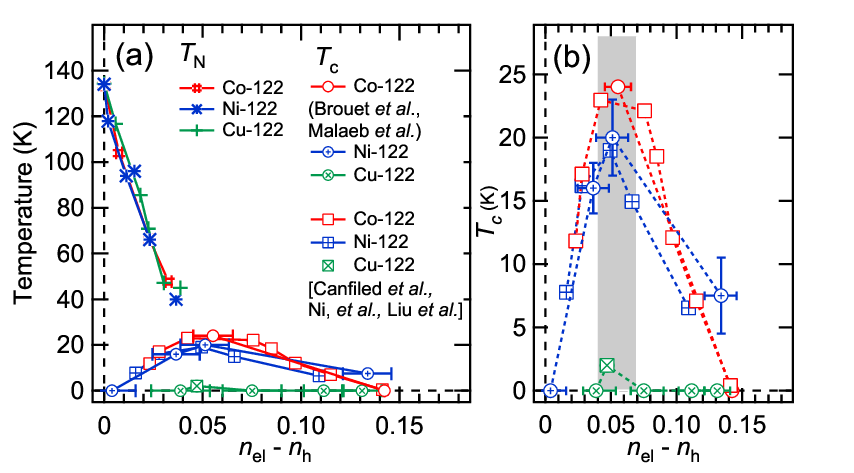}
\caption{$T_{c}$ and $T_{\rm{N}}$ of Ba(Fe$_{1-x}$$T$$_{x}$)$_2$As$_2$ ($T$ = Co, Ni, Cu) plotted as functions of the total FS volume $n_{\rm{el}}$ - $n_{\rm{h}}$ evaluated by ARPES. The data for Co-122 are taken from \cite{Canfield, Ni}. (b) is a magnified plot of $T_c$. The $T_c$ values for Co-, Ni-, and Cu-122 are taken from \cite{Canfield, Ni, Brouet, Malaeb, CLiu}. The $n_{\rm{el}}-n_{\rm{h}}$ of Co-, Ni-, and Cu-122 for small $x$'s has been estimated by linear interpolation between the lowest $n_{\rm{el}}-n_{\rm{h}}$ data and $n_{\rm{el}}-n_{\rm{h}}$ = 0. $T_{\rm{N}}$ for Ni-122 with $x$ = 0.02 and Cu-122 with $x$ = 0.04 are also plotted.} 
\label{sum2}
  \end{center}
\end{figure}

Now, we estimate the total hole and electron FS volumes in order to discuss to what extent the electron-doped system can be understood within the rigid-band model. First, let us estimate the volume of each hole and electron FS from the result of the FS mapping shown in Fig. \ref{FigTMBa122_4}. For the \textquotedblleft rugby\ ball\textquotedblright\ like hole FSs of Ni-122, we have estimated the volumes assuming two degenerate inner ($x$ = 0.02, 0.0375, and 0.05) and one outer ($x$ = 0.02, 0.0375, 0.05, and 0.08) hole FSs \cite{Brouet} (see also Supplemental Material in Fig. \ref{FigS3}). For the hole FS of Cu-122 with $x$ = 0.04, we applied the same method. For the electron FSs, we have also estimated the volumes assuming that two FSs are nearly degenerate as in the case of Co-122 \cite{1}. The total volume of the hole FSs, $n_{\rm{h}}$, that of the electron FSs, $n_{\rm{el}}$, and the total FS volume, namely, the difference $n_{\rm{el}}-n_{\rm{h}}$ are plotted as functions of nominal extra electron number in Fig. \ref{FigTMBa122_5}. In order to compare the present ARPES result with the previous studies of Co-122 \cite{Brouet, Malaeb}, $n_{\rm{h}}$ and $n_{\rm{el}}$ are plotted as functions of the nominal electron number, $x$ for Co, 2$x$ for Ni, and 3$x$ for Cu in Figs. \ref{FigTMBa122_5}(a) and \ref{FigTMBa122_5}(b). (Data for Co-122 with $\it{x}<$ 0.05 are not plotted here because of the antiferromagnetic band folding and hence the difficulty in estimating $\it{n_{\rm{h}}}$ and $\it{n_{\rm{el}}}$.) In the previous ARPES result on Co-122 \cite{Brouet, Malaeb}, the plots of $n_{\rm{h}}$ from the non-doped to overdoped samples decreases  gradually with increasing Co concentration as shown in Fig. \ref{FigTMBa122_5}(a). The $n_{\rm{h}}$ of Ni-122 exhibits the same behavior. In Fig. \ref{FigTMBa122_5}(c), one finds that $n_{\rm{el}}-n_{\rm{h}}$ for Co-122 \cite{Brouet} is proportional to $x$  for $x\geq$ 0.07, where the anti-ferromagnetic order disappears. On the other hand, the value of $n_{\rm{el}}-n_{\rm{h}}$ for Ni-122 increases, but seems to be slightly deviated downwards from the linear relationship predicted by the rigid-band model, indicating that signature of the deviation from the rigid-band model starts to be seen from Ni-122. As for Cu-122, $n_{\rm{el}}-n_{\rm{h}}$ is even smaller than that of Ni-122 in the entire doping range. The present ARPES result suggests that although most of the doped electrons participate in the formation of FSs, part of them may be partially localized and do not fill the energy bands as predicted by the rigid-band model. Recent Neutron scattering study has also reported that the electron counting based on the rigid-band model is inappropriate \cite{MGKim}.

The plots of $n_{\rm{h}}$ for Co-, Ni-, and Cu-122 shown in Fig. 4(a) almost coincide with each other. On the other hand, the $n_{\rm{el}}$ curve shown in Fig. \ref{FigTMBa122_5}(b) is shifted toward higher electron concentrations in going from Co, Ni, to Cu. This means that in going from Co-, Ni-, to Cu-122, the hole bands are shifted away from $E_{\rm{F}}$ by the same ratio, and that the relative positions of the electron and hole bands change (see Supplemental Material, Figs. S4 and S5 of \cite{supple}), indicating that the non-rigid band effect clearly exists.

So far, we have shown that electrons are indeed doped in the transition metal-substituted BaFe$_2$As$_2$, but the deviation from the rigid-band model exists in the band structure and the FS volumes and increases with increasing strength of the impurity potential. As shown in Fig. \ref{FigTMBa122_1}(b), the extra doped electron concentration determines the $\it{T_c}$ better than the impurity concentration, but the $T_{c,max}$ for Co-, Ni-, and Cu-122 are still significantly deviated from  each other in this plot. As for $T_{\rm{N}}$, the same plot fails, particularly in Cu-122. On the other hand, if the $T_c$ and $T_{\rm{N}}$ for Co-, Ni-, and Cu-122 are plotted as functions of $n_{\rm{el}}-n_{\rm{h}}$ deduced from the ARPES data \cite{footnote}, in Fig. \ref{sum2} the $T_{\rm{N}}$ for Co-, Ni-, and Cu-122 coincides with each other, and the $T_{c}$ maximum occurs at nearly the same $n_{\rm{el}}-n_{\rm{h}}$ value of $\sim$ 0.06 for all the systems. This means that the $T_{\rm{N}}$ values as well as the optimal composition of $\it{T_c}$ in the electron-doped BaFe$_2$As$_2$ are controlled by $n_{\rm{el}}-n_{\rm{h}}$ relatively well. Since the $\it{T_{c,max}}$ varies between the systems, the absolute $T_c$ values should be controlled not only by $n_{\rm{el}}-n_{\rm{h}}$, but also by other parameter(s) such as $n_{\rm{h}}$, $n_{\rm{el}}$, or the band structure. Since the energy position of the impurity 3$d$ level is lowered in going from Co, Ni, to Cu, it is natural to consider that the strong impurity potential suppresses the $\it{T_c}$ value.

Recently, Haverkort $\it{et\ al}$. \cite{Haverkort} and Berlijn $\it{et\ al}$. \cite{Berlijn} have studied how the electronic structure is affected by impurity potential when host atoms are substituted by other atoms. It has been shown that the hole-FS volume increases with increasing impurity potential \cite{Konbu}. For strong impurity potentials, substitution creates an impurity band split-off below the original host band \cite{Wadati, Nakamura, LZhang}. The impurity band removes electrons from the host band, leading to an apparent decrease in the electron occupation. This calculated result can be compared with the FS volume changes in the series $\it{T}$ = Co, Ni, and Cu for the same number of extra electrons estimated in the present ARPES study. In going from Ni to Cu, the impurity potential becomes stronger as demonstrated in Fig. \ref{Valence} and the hole-FS volume decreases. However, since the total electron number estimated from the FS volumes show a smaller value than that expected from the rigid-band model, both the calculation and experimental results show decrease in the electron occupation. For Co-122, the electronic structure almost follows the rigid-band model, because the Co 3$\it{d}$ level is near the Fe 3$\it{d}$ level.

By computing the configuration-averaged spectral function of disordered Co/Zn substituted BaFe$_2$As$_2$, Berlijn $\it{et\ al}$. \cite{Berlijn} have estimated the hole and electron volumes. For Co-122, the FS volume of Co-122 follows the Luttinger theorem, that is, the system follows the rigid-band model. On the other hand, for Zn-122, the Luttinger theorem breaks down. The result also indicates that the FS volume is smaller than that expected from the rigid-band model. Therefore, the increased deviation from the rigid-band model becomes stronger in going from Co, Ni, to Cu-122 is consistent with the calculated result for Zn-122. Thus, to understand the effect of the transition-metal atom substitution for Fe atoms on the electronic structure, the strength of the impurity potential should be taken into account. The calculations have also shown that all the spectral features of Zn-122 are broadened compared to Co-122. In fact, the hole and electron bands for Ni- and Cu-122 are not clearly resolved into multiple bands, whereas those of Co-122 and BaFe$_2$As$_2$ have been clearly resolved. In the studies of conventional binary random alloys, photoemission spectroscopy and theoretical calculation by the coherence potential approximation (CPA) \cite{Kirkpatrick, Rao, Winter, Punkkinen, Hasegawa, Hufner} have shown that the rigid-band model breaks down when the potential difference between two constituent atoms is comparable to or exceeds the band widths. In the latter case of alloys, characteristic features in the densities of states of the two constituent materials survive. In this sense, the present ARPES result may be understood in analogy of the conventional binary alloys.

In conclusion, we have performed an ARPES study of the transition metal-substituted Ba(Fe$_{1-x}T_x$)$_2$As$_2$ ($T$ = Ni, Cu) to investigate the validity of and the deviation from the rigid-band model, which has been usually assumed for Co-122. We find that, although the rigid-band model works qualitatively and electrons are indeed doped, the electron FS volumes in Ni- and Cu-122 are smaller than the value which is expected from the rigid-band model. This result suggests that part of electrons doped by substitution of Ni or Cu preferentially occupy the Ni 3$d$ or Cu 3$d$ states, or are trapped around the impurity sites, and do not behave like a mobile carrier. We found that the $T_{\rm{N}}$ in the electron-doped BaFe$_2$As$_2$ is determined by $n_{\rm{el}}-n_{\rm{h}}$, while $T_c$ is determined by both $n_{\rm{el}}-n_{\rm{h}}$, the band structure, and chemical disorder, which are affected by the impurity potential. 

Enlightening discussions with H. Wadati, G. A. Sawatzky, R. M. Fernandes, and M. Ogata are gratefully acknowledged. ARPES experiments were carried out at KEK-PF (Proposal No. 2009S2-005). This work was supported by an A3 Foresight Program from Japan Society for the Promotion of Science, a Grant-in-Aid for Scientific Research on Innovative Area \textquotedblleft{Materials Design through Computics: Complex Correlation and Non-Equilibrium Dynamics}\textquotedblright, and a Global COE Program \textquotedblleft Physical Sciences Frontier\textquotedblright, MEXT, Japan. S.I. acknowledges support from the Japan Society for the Promotion of Science for Young Scientists.

\section{Supplemental Material}

\subsection{1. Fermi surface mapping for Ni-122 and Cu-122}
\begin{figure*}[t]
  \begin{center}
\leavevmode
    \includegraphics[width=15cm]{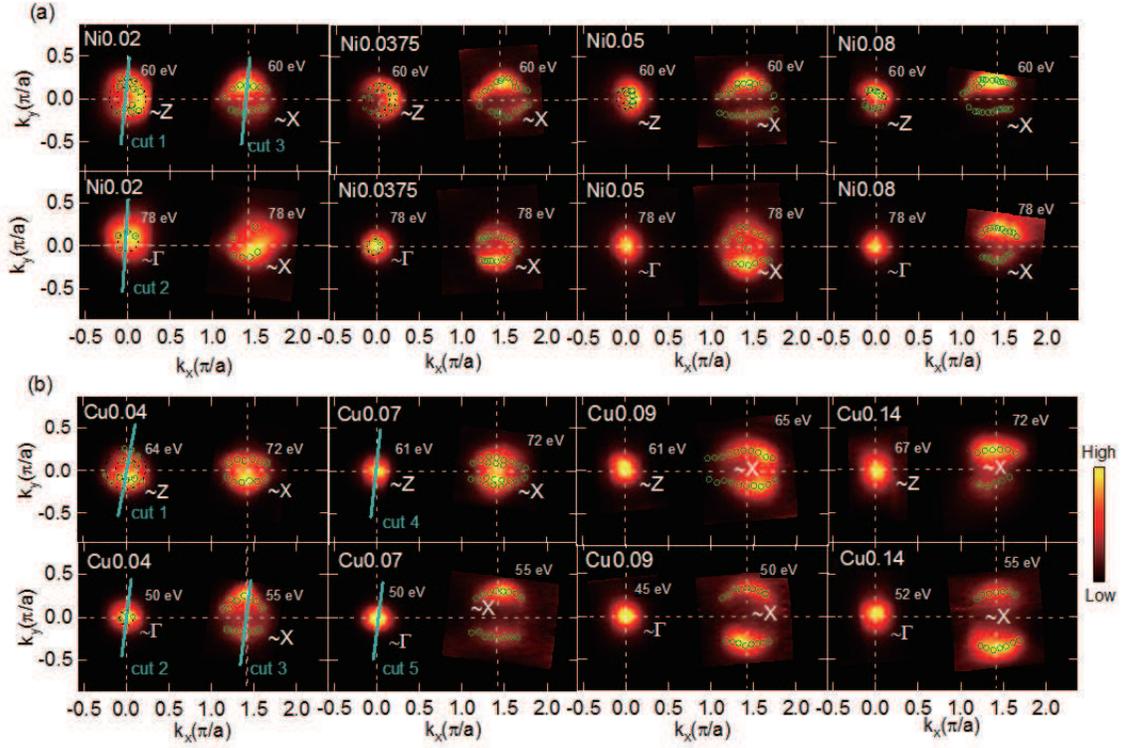}
\caption{ARPES intensity plot of Ba(Fe$_{1-x}$Ni$_x$)$_2$As$_2$ (Ni-122) and Ba(Fe$_{1-x}$Cu$_x$)$_2$As$_2$ (Cu-122). Intensity has been integrated over the energy range $\pm$ 5 meV of $E_{\rm{F}}$. Solid green and dashed black circles on the Fermi surfaces (FSs) show experimentally determined Fermi momentum ($k_{\rm{F}}$) and schematic FSs, respectively. (a): Ni-122 with $x$ = 0.02, 0.0375, 0.05, and 0.08 referred to Ni0.02, Ni0.0375, Ni0.05, and Ni0.08, respectively. $k_z$s measured with photon energies of $h\mu$= 60 and 78 eV are near the $\sim$ Z and $\sim \Gamma$ points, respectively. (b): Cu-122 with $x$ = 0.04, 0.07, 0.09, and 0.14 referred to Cu0.04, Cu0.07, Cu0.09, and Cu0.14, respectively.} 
\label{FigS1}
  \end{center}
\end{figure*}

Figure \ref{FigS1}(a) shows ARPES intensity plots of Ba(Fe$_{1-x}$Ni$_x$)$_2$As$_2$ (Ni-122) in@two-dimensional momentum space for the underdoped ($x$ = 0.02), slightly underdoped ($x$ = 0.0375), optimally doped ($x$ = 0.05), and overdoped ($x$ = 0.08) samples. As for the hole FSs, the $k_x$-$k_y$ cross-sectional area of one of the hole Fermi surfaces (FSs) around the Z point ($k_z \sim2\pi/c$) decreases with increasing $x$ as shown in the upper panel of Fig. \ref{FigS1}(a), and that around the $\Gamma$ point ($k_z \sim$ 0) disappears as shown in the lower panel of Fig. \ref{FigS1}(a) in the optimally-doped and slightly underdoped Ni-122. This indicates that the hole FS of Ni-122 has strong three dimensionality as in the case of the parent \cite{Malaeb}. On the other hand, the area of the electron FSs around the X point as shown in Fig. \ref{FigS1}(a) gradually expands with electron doping. In Fig. \ref{FigS1}(b), ARPES intensity plot of Ba(Fe$_{1-x}$Cu$_x$)$_2$As$_2$ (Cu-122) is shown for $x$ = 0.04, 0.07, 0.09, and 0.14. The hole FS has been observed at Cu-122 with $x$ =0.04 at the Z and $\Gamma$ points, while the hole FSs are not clearly observed even though there are finite intensities at and around the $\Gamma$ and Z points for $x$ = 0.07, 0.09, and 0.14 (see Fig. \ref{FigS2}). On the other hand, one can see a large electron FS around the X point for all the samples. The band-structure calculation of the parent compound BaFe$_2$As$_2$ \cite{Singh} suggests that the hole FS should exist at the Z point if the rigid-band model is applicable. However, the hole band of Cu-122 does not cross the Fermi level ($E_{\rm{F}}$) even around the Z point as shown in Fig. \ref{FigS2}, indicating that the hole FSs disappear for x = 0.07, 0.09 and 0.14.

\subsection{2. Band dispersions}

\begin{figure*}[t]
  \begin{center}
\leavevmode
    \includegraphics[width=15cm]{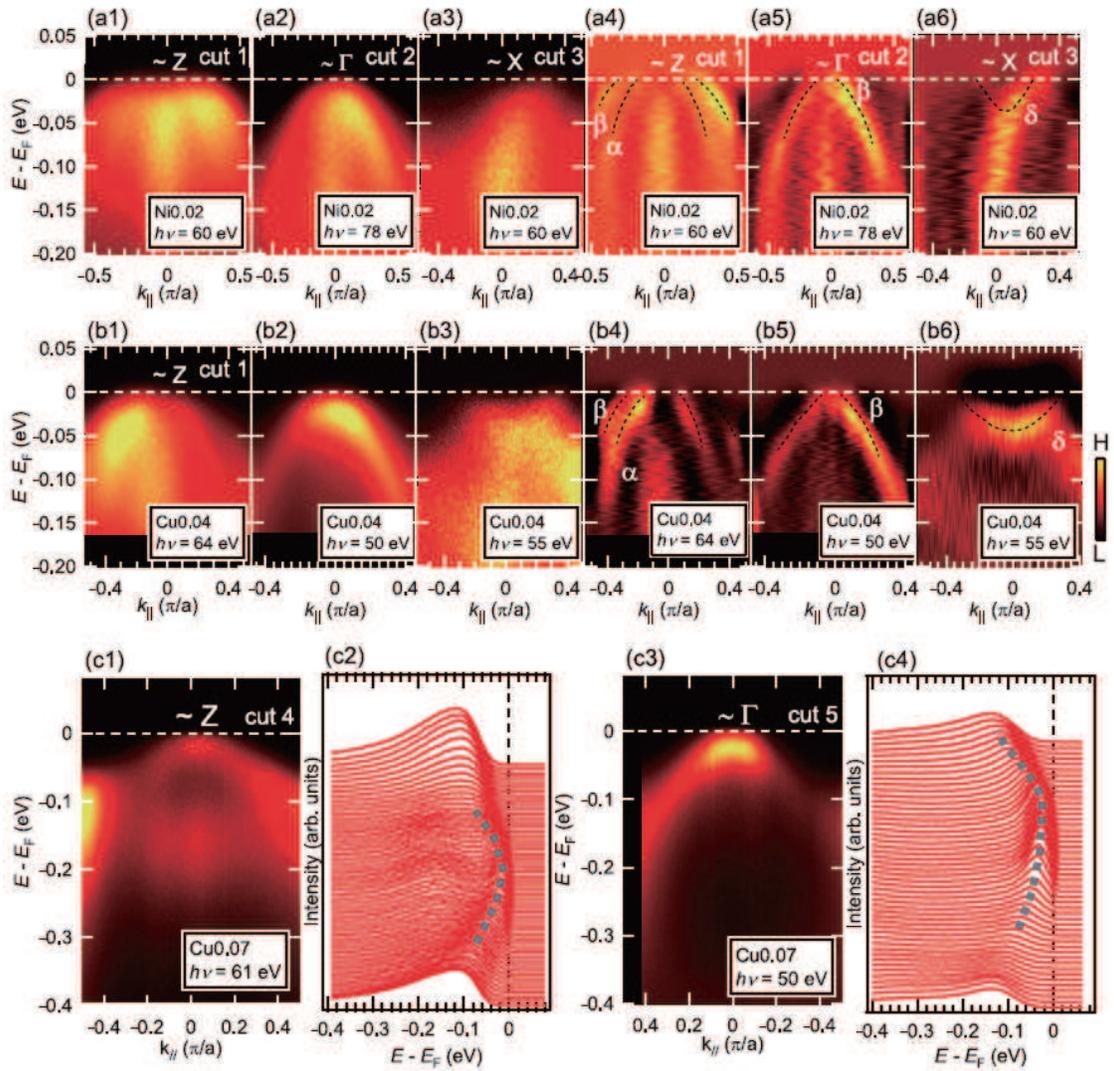}
\caption{ARPES spectra of Ni-122 and Cu-122. (a1)-(a3): Energy-momentum ($E-k$) plot around the Z, $\Gamma$ and X points for Ni0.02. (a4)-(a6): Second-derivative $E-k$ plots of (a1)-(a3). (b1)-(b3): $E-k$ plot around the Z, $\Gamma$ and X points for Cu0.04. (b4)-(b6): Second-derivative $E-k$ plots of (b1)-(b3). (c1),(c3): $E-k$ plots around the Z and $\Gamma$ points for Cu0.07. Even around the Z point, the band does not cross the $E_{\rm{F}}$. Panels (c2) and (c4) are EDCs corresponding to panels (c1) and (c3), respectively. Gray dashed curves are guide to the eye.} 
\label{FigS2}
  \end{center}
\end{figure*}

The energy-momentum ($E-k$) intensity plots of the hole and electron bands corresponding to cuts 1-3 for Ni-122 and cuts 1-5 for Cu-122 in Fig. \ref{FigS1} are shown in Figs. \ref{FigS2}(a1)-(a3) and Figs. \ref{FigS2}(b1)-(b3) and (c1)-(c4), respectively. We refer to the inner and outer hole bands as $\alpha$ and $\beta$, respectively, while we refer to the electron band as $\delta$. Concerning the electron bands around the X point of Ni-122 and Cu-122, one cannot resolve the two electron bands. The hole bands denoted by $\alpha$+$\beta$ of Cu-122 with $x$ = 0.07 does not cross $E_{\rm{F}}$ around the Z and $\Gamma$ points as shown in Figs. \ref{FigS2}(c1)-(c4).

\begin{figure*}[t]
  \begin{center}
\leavevmode
    \includegraphics[width=15cm]{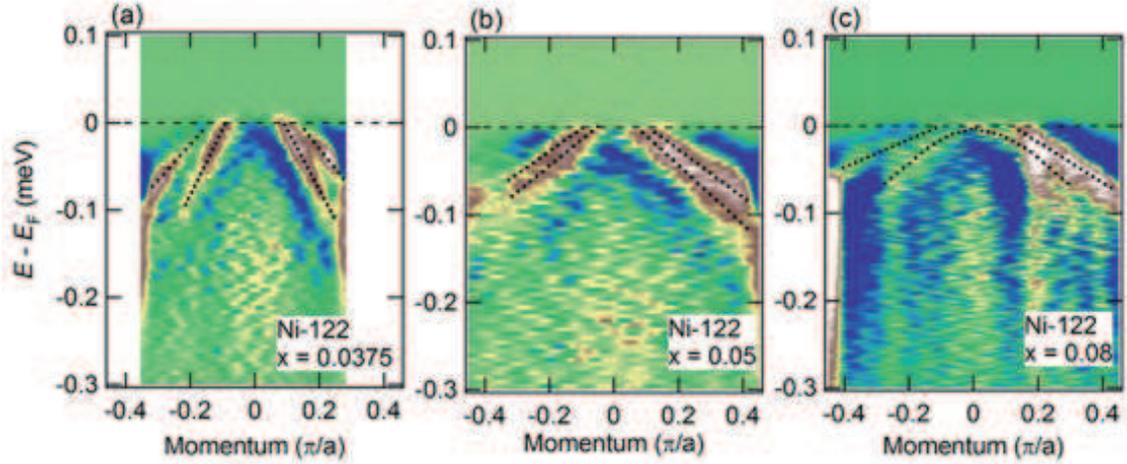}
\caption{(a)-(c): Second-derivative $E-k$ plots of Ni-122 taken at $h\nu$ = 60 eV, corresponding to $k_z$ around the Z point. Dotted lines are guide to the eye to trace hole bands.} 
\label{FigS3}
  \end{center}
\end{figure*}

In order to show how we have deduced the number of hole bands in Ni-122, we show the second-derivative of the EDCs in $E-k$ space. As shown in Fig. \ref{FigS3}, in the slightly undedoped ($x$ = 0.0375) and optimally doped ($x$ = 0.05) samples, two hole bands cross the $E_{\rm{F}}$, while in the overdoped ($x$ = 0.08) sample, one of them is slightly below the $E_{\rm{F}}$. For the estimation of the hole FS volume, we assume one outer hole FS for $x$ = 0.02, 0.0375, 0.05, and 0.08 samples and two degenerate inner hole FSs for $x$ = 0.02, 0.0375, and $x$ = 0.05 samples \cite{Brouet}.

\subsection{3. Shifts of hole- and electron-energy bands}

\begin{figure*}[t]
  \begin{center}
\leavevmode
    \includegraphics[width=15cm]{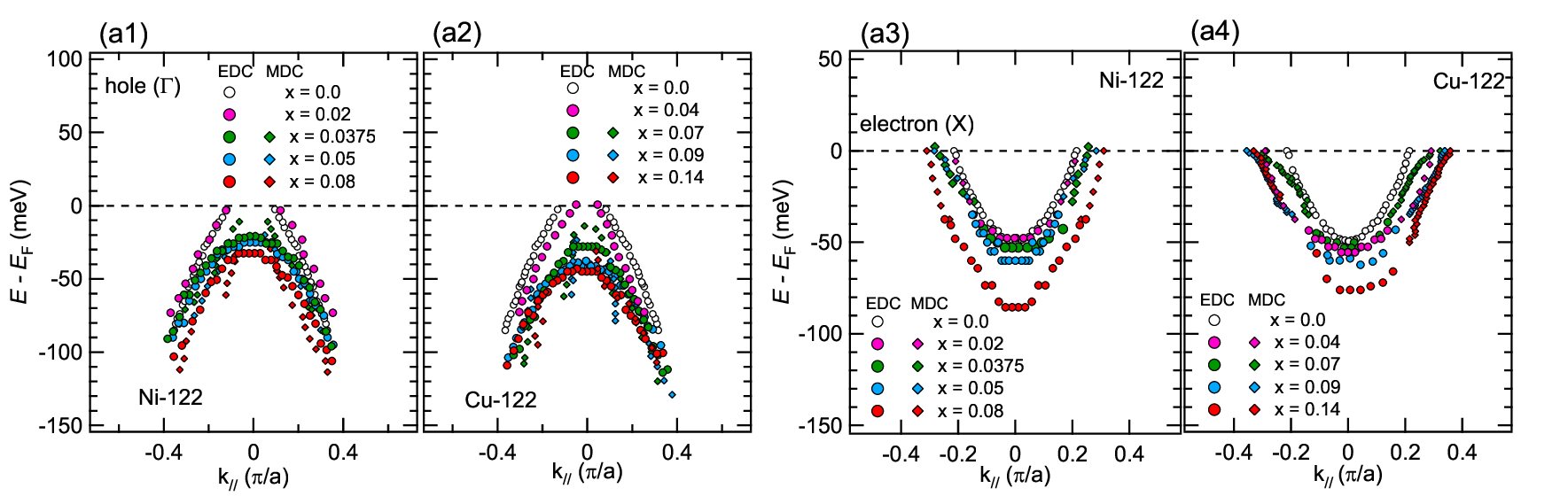}
\caption{Hole bands around the $\Gamma$ point [(a1) and (a2)] and the electron bands around the X point [(a3) and (a4)] for Ni-122 and Cu-122. Band dispersions of BaFe$_2$As$_2$ are taken from \cite{Malaeb}.} 
\label{FigS4}
  \end{center}
\end{figure*}

As shown in Figs. \ref{FigS4}(a1) and \ref{FigS4}(a2), the hole band maxima of Ni-122 and Cu-122 around the $\Gamma$ point are shifted by $\sim$ 30 meV and $\sim$ 40 meV, respectively, compared with that of the band dispersions of BaFe$_2$As$_2$ \cite{Malaeb}. On the other hand, the shifts of the minimum energies of the electron bands of Ni-122 and Cu-122 are $\sim$ 35 meV and $\sim$ 25 meV, respectively, compared with that of the band dispersions of BaFe$_2$As$_2$ \cite{Malaeb} as shown in Figs. \ref{FigS4}(a3) and \ref{FigS4}(a4). The energy shifts of the hole and electron bands for Ni-122 are almost the same, while those for Cu-122 show a remarkable difference, indicating that the deviation from the rigid-band model becomes strong in going from Ni- to Cu-122.

In Figs. S5(a) and S5(b), we have plotted the doping dependences of the energies of the hole-band maximum and the electron-band minimum of Fig. \ref{FigS4} together with the ARPES results on Co-122 \cite{Brouet, CLiu, Neupane}. For the hole band, Co- and Ni-122 show the almost rigid-band shift. On the other hand, the hole band of Cu-122 is strongly deviated from the rigid-band model. The energy shifts of the electron band for Ni- and Cu-122 show a deviation from the rigid-band model and those of Co-122. These non-rigid-band shifts are consistent with the results of the $n_{\rm{h}}$ and $n_{\rm{el}}$ curves shown in Figs. 4(a) and 4(b) in the main text.

\begin{figure*}[t]
  \begin{center}
\leavevmode
    \includegraphics[width=15cm]{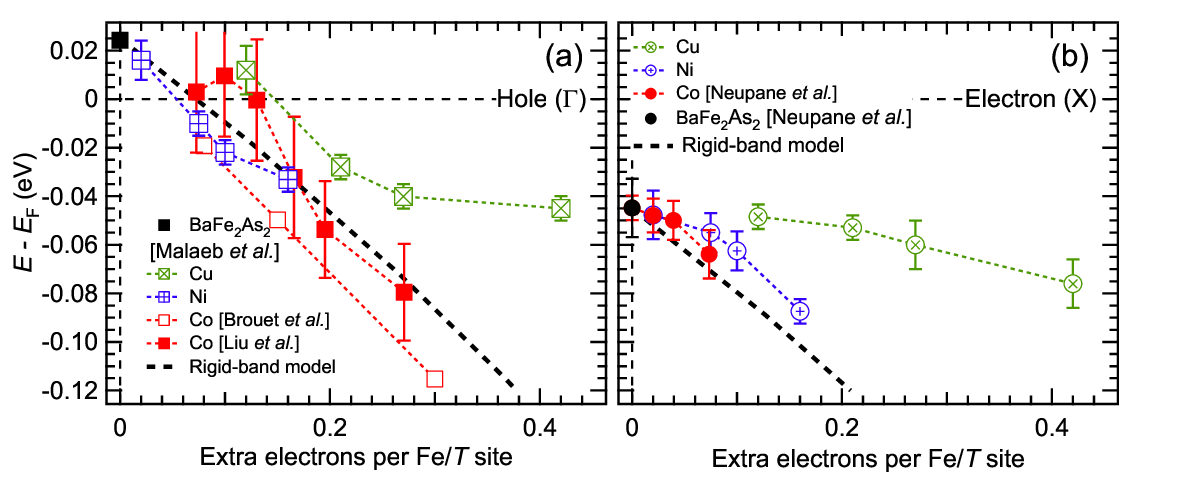}
\caption{Shifts of the hole-band maximum around the $\Gamma$ point (a) and the electron-band minimum around the X point (b). The dashed lines show the shifts expected from the rigid-band model based on the non-magnetic band structure of BaFe$_2$As$_2$ \cite{Singh}. Data of BaFe$_2$As$_2$ and Co-122 for the hole band are taken from \cite{Malaeb, Brouet, CLiu} and that of BaFe$_2$As$_2$ and Co-122 for the electron band are taken from \cite{Neupane}.} 
\label{FigS5}
  \end{center}
\end{figure*}


\begin{thebibliography}{99}


%\bibitem{Kamihara2}Y. Kamihara, T. Watanabe, M. Hirano, and  H. Hosono, 
%J. Am. Chem. Soc. \textbf{130}, 3296 (2008).

%\bibitem{Uchida} S. Uchida, J. Phys. Soc. Jpn. \textbf{77} Suppl. C, 9 (2008).

%\bibitem{Day} C. Day, Phys. Today \textbf{61}, 11 (2008).


\bibitem{Rotter} M. Rotter, M. Tegel, and D. Johrendt, 
Phys. Rev. Lett. \textbf{101}, 107006 (2008).

\bibitem{Sasmal} K. Sasmal, B. Lv, B. Lorenz, A. M. Guloy, F. Chen, Y.-Y. Xue, and C.-W. Chu, Phys. Rev. Lett. \textbf{101}, 107007 (2008).

\bibitem{Sefat} A. S. Sefat, R. Jin, M. A. McGuire, B. C. Sales, D. J. Singh, and D. Mandrus, Phys. Rev. Lett. \textbf{101}, 117004 (2008).

\bibitem{Canfield} P. C. Canfield, S. L. Bud'ko, Ni Ni, J. Q. Yan, and A. Kracher, Phys. Rev. B \textbf{80}, 060501(R) (2009).

\bibitem{Ni} N. Ni, A. Thaler, J. Q. Yan, A. Kracher, E. Colombier, S. L. Bud'ko, P. C. Canfield, and S.T. Hannahs, Phys. Rev. B \textbf{82}, 024519 (2010).
\bibitem{Li} L. J. Li, Y. K. Luo, Q. B. Wang, H. Chen, Z. Ren, Q. Tao, Y. K. Li,X. Lin, M. He, Z. W. Zhu, G. H. Cao, and Z. A. Xu, New J. Phys. \textbf{11}, 025008 (2009).
\bibitem{Fukuzumi} Y. Fukuzumi, K. Mizuhashi, K. Takenaka, and S. Uchida, Phys. Rev. Lett. \textbf{76}, 684 (1996).




\bibitem{Ding} H. Ding, P. Richard, K. Nakayama, K. Sugawara, T. Arakane, Y. Sekiba, A. Takayama, S. Souma, T. Sato, T. Takahashi, Z. Wang, X. Dai, Z. Fang, G. F. Chen, J. L. Luo, and N. L. Wang, Europhys. Lett. \textbf{83}, 47001 (2008).

\bibitem{Wray}L. Wray, D. Qian, D. Hsieh, Y. Xia, L. Li, J. G. Checkelsky, A. Pasupathy, K. K. Gomes, C. V. Parker, A. V. Fedorov, G. F. Chen, J. L. Luo, A. Yazdani, N. P. Ong, N. L. Wang, and M. Z. Hasan, 
Phys. Rev. B \textbf{78}, 184508 (2008).

\bibitem{Sato1} T. Sato, K. Nakayama, Y. Sekiba, P. Richard, Y.-M. Xu, S. Souma, T. Takahashi, G. F. Chen, J. L. Luo, N. L. Wang, and H. Ding, Phys. Rev. Lett. \textbf{103}, 047002 (2009).

\bibitem{Malaeb}W. Malaeb, T. Yoshida, A. Fujimori, M. Kubota, K. Ono, K. Kihou, P. M. Shirage, H. Kito, A. Iyo, H. Eisaki, Y. Nnakajima, T. Tamegai, and R. Ariata, J. Phys. Soc. Jpn. \textbf{78}, 123706 (2009).
\bibitem{Brouet} V. Brouet, M. Marsi, B. Mansart, A. Nicolaou, A. Taleb-Ibrahimi, P. LeFevre, F. Bertran, F. Rullier-Albenque, A. Forget, and D. Colson, Phys. Rev. B \textbf{80}, 165115 (2009).

\bibitem{Neupane} M. Neupane, P. Richard, Y.-M. Xu, K. Nakayama, T. Sato, T. Takahashi, A. V. Federov, G. Xu, X. Dai, Z. Fang, Z. Wang, G.-F. Chen, N.-L. Wang, H.-H. Wen, and H. Ding, Phys. Rev. B $\textbf{83}$, 094522 (2011). 

\bibitem{Sekiba} Y. Sekiba, T. Sato, K. Nakayama, K. Terashima, P. Richard, J. H. Bowen, H. Ding, Y.-M. Xu, L. J. Li, G. H. Cao, Z.-A. Xu, and T. Takahashi, New J. Phys. \textbf{11}, 025020 (2009).

\bibitem{Terashima} K. Terashima, Y. Sekiba, J. H. Bowens, K. Nakayama, T. Kawahara, T. Sato, P. Richard, Y.-M. Xu, L. J. Li, G. H. Cao, Z.-A. Xu, H. Ding, and T. Takahashi, PNAS \textbf{106}, 7333 (2009).

\bibitem{Nakayama} K. Nakayama, T. Sato, P. Richard, Y.-M. Xu, Y. Sekiba, S. Souma, G. F. Chen, J. L. Luo, N. L. Wang, H. Ding, T. Takahashi, 
Europhys. Lett. \textbf{85}, 67002 (2009).
 


\bibitem{Chang Liu} C. Liu, A. D. Palczewski, R. S. Dhaka, Takeshi Kondo, R. M. Fernandes, E. D. Mun, H. Hodovanets, A. N. Thaler, J. Schmalian, S. L. Bud'ko, P. C. Canfield, and A. Kaminski, Phys. Rev. B \textbf{84}, 020509(R) (2011).

\bibitem{Vilmercati} P. Vilmercati, A. Fedorov, I. Vobornik, U. Manju, G. Panaccione, A. Goldoni, A. S. Sefat, M. A. McGuire, B. C. Sales, R. Jin, D. Mandrus, D. J. Singh, and N. Mannella, Phys. Rev. B \textbf{79}, 220503(R) (2009). 



\bibitem{Wadati} H. Wadati, I. Elfimov, and G. A. Sawatzky, Phys. Rev. Lett. \textbf{105}, 157004 (2010).

\bibitem{Konbu} S. Konbu, K. Nakamura, H. Ikeda, and R. Arita, J. Phys. Soc. Jpn. \textbf{80}, 123701 (2011).





\bibitem{Singh} D. J. Singh, Phys. Rev. B \textbf{78}, 094511 (2008).

\bibitem{Bianconi} A. Bianconi, J. Supercond. Nov. Magn. \textbf{18}, 625 (2005); D. Innocenti, N. Poccia, A. Ricci, A. Valletta, S. Caprara, A. Perali, and A. Bianconi, Phys. Rev. B \textbf{82}, 184528 (2010); D. Innocenti, S. Caprara, N. Poccia, A. Ricci, A. Valletta, and A. Bianconi, Supercond. Sci. Technol. \textbf{24}, 015012 (2011).
\bibitem{supple} See Supplemental Material at [URL will be inserted by publisher] for details of Fermi surface mapping in the $k_x$ - $k_y$ plane and the hole and electron bands. 



\bibitem{1} For the slightly underdoped Ni-122 with $x$ = 0.0375, a band folding has not been observed, probably because the $T_{\rm{N}}$ is relatively low. Therefore, it is reasonable to estimate the $n_{\rm{h}}$ and $n_{\rm{el}}$ from the data below $T_{\rm{N}}\sim$ 40 K. In Cu-122, two electron FSs have been observed at some photon energies, but not all, and therefore, we have taken into account the two FS sheets to estimate the number of electron carriers. 

%\bibitem{Olariu} A. Olariu $\it{et\ al.}$, %F. Rullier-Albenque, D. Colson, and A. Forget, 
%Phys. Rev. B \textbf{83}, 054518 (2011).

\bibitem{CLiu} C. Liu, A. D. Palczewski, R. S. Dhaka, Takeshi Kondo, R. M. Fernandes, E. D. Mun, H. Hodovanets, A. N. Thaler, J. Schmalian, S. L. Bud'ko, P. C. Canfield, and A. Kaminski, Phys. Rev. B \textbf{84}, 020509(R) (2011).

\bibitem{MGKim} M. G. Kim, J. Lamsal, T. W. Heitmann, G. S. Tucker, D. K. Pratt, S. N. Khan, Y. B. Lee, A. Alam, A. Thaler, N. Ni, S. Ran, S. L. Bud'ko, K. J. Marty, M. D. Lumsden, P. C. Canfield, B. N. Harmon, D. D. Johnson, A. Kreyssig, R. J. McQueeney, and A. I. Goldman, Phys. Rev. Lett. \textbf{109}, 167003 (2012).
\bibitem{footnote} $n_{\rm{el}}-n_{\rm{h}}$ has been interpolated linearly between the lowest experimental value of $n_{\rm{el}}-n_{\rm{h}}$ and $n_{\rm{el}}-n_{\rm{h}}$ = 0. 

\bibitem{Haverkort} M. W. Haverkort, I. S. Elfimov, and G. A. Sawatzky,
arXiv:1109.4036.


\bibitem{Berlijn} T. Berlijn, C.-H. Lin, W. Garber, and W. Ku, Phys. Rev. Lett. \textbf{108}, 207003 (2012).

\bibitem{Nakamura} K. Nakamura, R. Arita, and H. Ikeda, Phys. Rev. B \textbf{83}, 144512 (2011).

\bibitem{LZhang} L. Zhang and D. J. Singh, Phys. Rev. B \textbf{80}, 214530 (2009).

\bibitem{Kirkpatrick} S. Kirkpatrick, B. Velicky, and H. Ehrenreich, Phys. Rev.
B \textbf{1}, 3250 (1970).

\bibitem{Rao} R. S. Rao, A. Bansil, H. Asonen, and M. Pessa, Phys.
Rev. B \textbf{29}, 1713 (1984).

\bibitem{Winter} H. Winter, P. J. Durham, W. M. Temmerman, and G. M. Stocks, Phys. Rev. B \textbf{33}, 2370 (1986).

\bibitem{Punkkinen} M. P. J. Punkkinen, K. Kokko, M. Ropo, I. J. Vayrynen, L. Vitos, B. Johansson, and J. Kollar, Phys. Rev. B \textbf{73}, 024426 (2006).

\bibitem{Hasegawa} H. Hasegawa and J. Kanamori, J. Phys. Soc. Jpn. \textbf{31},
382 (1971).

\bibitem{Hufner} S. H$\rm{\ddot{u}}$fner, G. K. Wertheim, R. L. Cohen, and J. H. Wernick, Phys. Rev. Lett. \textbf{28}, 488 (1972).



\end{thebibliography}
\end{document}